\documentclass[10pt]{article}
\pdfoutput=1
\usepackage{amssymb,amsmath,mathrsfs,enumerate}
\usepackage{graphicx,rotate,multicol}
\usepackage[margin=10pt,labelfont=bf]{caption}
\usepackage{cite}
\usepackage[colorlinks=true,
            linkcolor=red,
            urlcolor=blue,
            citecolor=blue]{hyperref}

\usepackage{color}
\usepackage{tikz,braket}
\usetikzlibrary{arrows,positioning, shapes.geometric}
\long\def\rpl#1!!#2!!{\textcolor{red}{#1} \textcolor{blue}{#2}}

\def\tb{\tan\beta}
\def\luv{\Lambda_{\rm UV}}
\def \order(#1){{\cal O} \left(#1 \right)}

\evensidemargin=\oddsidemargin
\def\Eqn#1{Eq.\ (\ref{#1})}
\def\Eqs#1#2{Eqs.\ (\ref{#1}) and (\ref{#2})}
\allowdisplaybreaks
\begin{document}

\begin{center}
	{\Large \bf Search for a `stable alignment limit' in two Higgs-doublet models } \\
	\vspace*{1cm} {\sf Dipankar
		Das$^{a,}$\footnote{d.das@saha.ac.in}, ~Ipsita Saha$^{b,}$\footnote{tpis@iacs.res.in}} \\
	\vspace{10pt} {\small } {\em $^a$Theory Physics Division, Saha Institute of Nuclear
		Physics, 1/AF Bidhan Nagar, Kolkata 700064, India \\
		$^b$Department of Theoretical Physics, 
		Indian Association for the Cultivation of Science, \\
		2A $\&$ 2B, Raja S.C. Mullick Road, Kolkata 700032, India}
	
	\normalsize
\end{center}

\begin{abstract}
	We study the conditions required to make the 2HDM scalar potential stable up to the Planck
	scale. The lightest CP-even scalar is assumed to have been found at the LHC and the {\em alignment limit} is imposed in view of the LHC Higgs data. We find that ensuring
	stability up to scales $\gtrsim 10^{10}$~GeV necessitates the introduction of a soft breaking
	parameter in the theory. Even then, some interesting correlations between the nonstandard masses
	and the soft breaking parameter need to be satisfied. Consequently, a 2HDM becomes completely
	determined by only two nonstandard parameters, namely, $\tb$ and a mass parameter, $m_0$, with
	$\tb \gtrsim 3$. These observations make a 2HDM, in the {\em stable alignment limit}, more
	predictive than ever.
\end{abstract}

\bigskip
\section{Introduction}\label{intro}
With the discovery of a Higgs-like particle at the Large Hadron Collider (LHC)  
\cite{Aad:2012tfa,Chatrchyan:2012ufa}, all the parameters of the Standard Model (SM) are now known
and the fate of the SM is sealed. Taking all the uncertainties of the current
experimental data into account, it has been found that the SM scalar potential becomes unstable 
somewhere within $10^8$-$10^{10}$~GeV \cite{Bezrukov:2012sa,Degrassi:2012ry,Buttazzo:2013uya,Branchina:2013jra}.
In fact, it has been shown that for the SM scalar potential to be stable
all the way up to the Planck scale $(M_P = 10^{19}~\rm GeV)$ the Higgs
mass $(m_h)$ needs to be in the following range  \cite{Buttazzo:2013uya}:
\begin{eqnarray}
m_h > 129.6~{\rm GeV} + 2.0\left(M_t[{\rm GeV}]-173.34~{\rm GeV}\right)-0.5~{\rm GeV}\left(\frac{\alpha_s(M_Z)-0.1184}{0.0007}\right) \pm 0.3~{\rm GeV} \,, 
\end{eqnarray}
where, $M_t$ denotes the top quark pole mass and $\alpha_s(M_Z)$ is the strong coupling
constant at the $Z$-boson mass scale. Hence, an SM Higgs boson with mass in the range 124-126 GeV
certainly disfavors the possibility of having an absolutely stable vacuum up to $M_P$.
As a way out of this vexing situation, it has been suggested that while absolute stability of the SM potential
might be a tall ask, a metastable vacuum is entirely consistent with the current
experimental value of the Higgs mass \cite{Degrassi:2012ry,Branchina:2014rva}. 
Nevertheless, the problem of vacuum stability 
in the SM remains one of the most discussed topics after the Higgs discovery and 
often has been taken as a hint for the intervention of some new physics. 

Our objective in this paper is to investigate whether the two-Higgs-doublet models (2HDMs) \cite{Branco:2011iw} can do better in this respect. 2HDMs extend the scalar potential of the SM by adding one extra scalar doublet and therefore, rank amongst the simplest of  beyond the Standard Model (BSM) constructions. For decades, 2HDMs have attracted a lot of attention because the minimal supersymmetry relies on a 2HDM scalar structure. Another attractive feature of 2HDMs is that the value of the electroweak $\rho$-parameter remains unity
at the tree level. But one ominous consequence of introducing one additional scalar
doublet is that now we will have two Yukawa matrices for each type of fermion and
diagonalization of the fermion mass matrix will not guarantee the simultaneous diagonalization of the Yukawa matrices. As a result, there will be flavor changing neutral currents(FCNC),
at the tree level, mediated by neutral scalars. 
This problem was addressed by Glashow and Weinberg \cite{Glashow:1976nt} and independently by Paschos\cite{Paschos:1976ay}. 
According to Glashow-Weinberg-Paschos theorem, tree level 
FCNC can be avoided altogether if suitable arrangements are made such that fermions 
of a particular charge receive their masses from a particular scalar 
doublet. Usually, a $Z_2$ symmetry, under which $\phi_1 \to \phi_1$ and $\phi_2 \to -\phi_2$, is employed to achieve this. Proper assignments of the $Z_2$ charge to different
fermions then complete the purpose. Labeling as $\phi_2$ the doublet that couples to the up-type quarks, the following four conventional variants of 2HDMs emerge from the $Z_2$ charge assignments to the fermions: 
\begin{itemize}
\item Type I: All quarks and leptons couple only to the doublet $\phi_2$; 
\item Type II: $\phi_2$ couples to the up-type quarks and $\phi_1$ couples to down-type quarks and charged leptons; 
\item Type X or lepton specific: All quarks couple to $\phi_2$, while $\phi_1$ couples to the charged leptons; 
\item Type Y or flipped: Up-type quarks and charged leptons couple to $\phi_2$ and all down-type quarks couple to $\phi_1$.
\end{itemize}
Currently the measured values of Higgs signal strengths into different decay channels are consistent with
the corresponding SM expectations. In anticipation that the data will continue to agree with the SM with increasing
accuracy in the upcoming experiments, those BSM scenarios which can deliver an SM-like Higgs in some
limiting case, will hold the upper hand in future survival race. It is quite well known that 2HDMs promise an 
{\em alignment limit} \cite{Gunion:2002zf,Craig:2012vn,Carena:2013ooa} when an SM-like Higgs can be recovered in the form of the lightest CP-even scalar
in the 2HDM particle spectrum. Although vacuum stability constraints in 2HDMs have been studied previously
both before \cite{Ferreira:2009jb} and after \cite{Chakrabarty:2014aya} the Higgs discovery, the distinct
implications of the alignment limit in this context have never been emphasized before. Motivated by the LHC Higgs data, in this paper we concentrate exclusively on the alignment limit and explore
the consequences of demanding the stability of the 2HDM potential all the way up to the GUT and Planck 
scales. In the process we have uncovered many new and interesting features. We have found that the requirement
of stability of the 2HDM scalar potential compels us to introduce a soft breaking parameter and at 
the same time, entails a strong correlation between the soft breaking parameter and the other nonstandard 
masses. Accordingly, in this `stable alignment limit' a 2HDM is completely determined by only two
nonstandard parameters: (i) $\tan\beta$ ($=v_2/v_1$) which is the ratio of the two vacuum expectation 
values (vevs) and, (ii) a mass parameter ($m_0$). How the requirement of high scale
stability of the 2HDM scalar potential in the alignment limit leads us to this intriguing conclusion constitutes the
central theme of our paper.

The paper is organized in the following way: in Section~\ref{model} we
describe the model and the various constraints that we use in our study.
We present our results in Section~\ref{results}. In
Section~\ref{conclusions} we summarize important findings and draw our conclusions.

\section{The scalar potential}\label{model} 
The general 2HDM potential with $Z_2$ symmetry under which $\phi_1 \to \phi_1$
and $\phi_2 \to -\phi_2$, is usually written as
\begin{eqnarray}
 V &=& m_{11}^2 \phi_1^\dagger\phi_1 +
m_{22}^2\phi_2^\dagger\phi_2 -\left(m_{12}^2 \phi_1^\dagger\phi_2
+{\rm h.c.} \right) +\frac{\lambda_1}{2} \left(\phi_1^\dagger\phi_1
\right)^2 +\frac{\lambda_2}{2} \left(\phi_2^\dagger\phi_2 \right)^2
\nonumber \\ 
&& +\lambda_3 \left(\phi_1^\dagger\phi_1 \right)
\left(\phi_2^\dagger\phi_2 \right) +\lambda_4 \left(\phi_1^\dagger\phi_2
\right) \left(\phi_2^\dagger\phi_1 \right) +\left\{
\frac{\lambda_5}{2}
\left(\phi_1^\dagger\phi_2 \right)^2 +{\rm h.c.}\right\} \,,
\label{potential_notation1}
\end{eqnarray}
where the term proportional to $m_{12}^2$ breaks the $Z_2$ symmetry softly. For simplicity, we assume that all the parameters of the potential are real so that CP symmetry is not explicitly broken in the scalar sector. To minimize the potential, we
express the doublets as
\begin{eqnarray}
\phi_i = \left(\begin{array}{c}
\chi_i^+ \\
\frac{1}{\sqrt{2}} (h_i + i\eta_i)
\end{array} \right) \,,
\label{phi}
\end{eqnarray}
and define the minimum as follows:
\begin{eqnarray}
\braket{\phi_i}_{\rm min} = \left(\begin{array}{c}
0 \\
\frac{v_i}{\sqrt{2}} 
\end{array} \right) \, \equiv 
\left(\begin{array}{c}
0 \\
\frac{(x_i + iy_i)}{\sqrt{2}} 
\end{array} \right) \,.
\label{phi_min}
\end{eqnarray}
Note that, although the potential of \Eqn{potential_notation1} is explicitly CP conserving, the vevs can, in general, be complex\cite{Gunion:2005ja,Ferreira:2010hy}. Now the minimization conditions will read as follows:
 \begin{subequations}
 \label{min}
 \begin{eqnarray}
\frac{\partial V}{\partial h_1}\Big|_{\rm min} &=& 2(m_{11}^2x_1-m_{12}^2x_2) +\lambda_1x_1 (x_1^2+y_1^2)
+x_1(\lambda_3+\lambda_4) (x_2^2+y_2^2) \nonumber \\ && +\lambda_5(x_1x_2^2+2x_2y_1y_2-x_1y_2^2) ~=~ 0 \,,
 \label{min1} \\
 \frac{\partial V}{\partial h_2}\Big|_{\rm min} &=& 2(m_{22}^2x_2-m_{12}^2x_1) +\lambda_2x_2 (x_2^2+y_2^2)
 +x_2(\lambda_3+\lambda_4) (x_1^2+y_1^2) \nonumber \\ && +\lambda_5(x_2x_1^2+2x_1y_1y_2-x_2y_1^2) ~=~ 0 \,,
  \label{min2} \\
\frac{\partial V}{\partial \eta_1}\Big|_{\rm min} &=& 2(m_{11}^2y_1-m_{12}^2y_2) +\lambda_1y_1 (x_1^2+y_1^2)
+y_1(\lambda_3+\lambda_4) (x_2^2+y_2^2) \nonumber \\ && +\lambda_5(y_1y_2^2+2x_1x_2y_2-x_2^2y_1) ~=~ 0 \,,
 \label{min3} \\
\frac{\partial V}{\partial \eta_2}\Big|_{\rm min} &=& 2(m_{22}^2y_2-m_{12}^2y_1) +\lambda_2y_2 (x_2^2+y_2^2)
+y_2(\lambda_3+\lambda_4) (x_1^2+y_1^2) \nonumber \\ && +\lambda_5(y_2y_1^2+2x_1x_2y_1-x_1^2y_2) ~=~ 0 \,.
 \label{min4}
 \end{eqnarray}
 \end{subequations}
 It can be easily checked that the choice $y_1=y_2=0$ which implies the vevs are real, satisfies \Eqs{min3}{min4} trivially.
 In this case, \Eqs{min1}{min2} will take the following simpler forms:
 \begin{subequations}
 \begin{eqnarray}
 \frac{\partial V}{\partial h_1}\Big|_{\rm min} &=& 2(m_{11}^2v_1-m_{12}^2v_2) +v_1\left\{\lambda_1v_1^2 +(\lambda_3+\lambda_4+\lambda_5)v_2^2 \right\} ~=~ 0 \,, \label{min11} \\
  \frac{\partial V}{\partial h_2}\Big|_{\rm min} &=& 2(m_{22}^2v_2-m_{12}^2v_1) +v_2\left\{\lambda_2v_2^2 +(\lambda_3+\lambda_4+\lambda_5)v_1^2 \right\} ~=~ 0 \,.
   \label{min22}
 \end{eqnarray}
 \end{subequations}
 Thus the assumption of real $v_1$ and $v_2$ is consistent with the minimization conditions of \Eqn{min} and in what
 follows, we shall work under this assumption.
 
When both the doublets receive vevs, the $Z_2$ symmetry is broken spontaneously too and we can rewrite
the 2HDM potential in the following form \cite{Gunion:1989we}:
\begin{eqnarray}
 V &=& 
 \beta_1 \left( \phi_1^\dagger\phi_1 - \frac{v_1^2}{2} \right)^2 
+\beta_2 \left( \phi_2^\dagger\phi_2 - \frac{v_2^2}{2} \right)^2 
+\beta_3 \left( \phi_1^\dagger\phi_1 + \phi_2^{\dagger}\phi_2 
- \frac{v_1^2+v_2^2}{2} \right)^2
\nonumber \\*
&&  
+ \beta_4 \left\{ 
(\phi_1^{\dagger}\phi_1) (\phi_2^{\dagger}\phi_2) -
(\phi_1^{\dagger}\phi_2) (\phi_2^{\dagger}\phi_1)
\right\}
+ \beta_5 \left( {\rm Re} ~ \phi_1^\dagger\phi_2 - \frac{v_1 v_2}{2} \right)^2 
+ \beta_6 \left( {\rm Im} ~ \phi_1^\dagger\phi_2 \right)^2 \,.
\label{potential_notation2}
\end{eqnarray}
In comparison with the previous case, in this notation $\beta_5$ plays
the role of the  soft breaking parameter. For easy understanding of the future results, we will often switch
between the parametrizations of \Eqs{potential_notation1}{potential_notation2}. Therefore, a comparison
between the parameters of \Eqs{potential_notation1}{potential_notation2} will be in order. It should be
emphasized that unlike \Eqn{potential_notation1}, \Eqn{potential_notation2} manifestly assumes that both
the doublets acquire vevs so that $m_{11}^2$ and $m_{22}^2$ in \Eqn{potential_notation1} can be traded
for $v_1$ and $v_2$ (using \Eqs{min11}{min22}) followed by suitable rearrangements of the quartic parameters to obtain \Eqn{potential_notation2}.
Since, in this paper, we restrict ourselves only to the {\em non-inert} cases where $\tan\beta$ is nonzero and
finite, above two parametrizations are equivalent to us. The connections between the two sets of parameters
are given by the following relations:
\begin{eqnarray}
m_{11}^2 =-(\beta_1v_1^2+\beta_3v^2)~;~ &&\lambda_1=2(\beta_1+\beta_3)~; \nonumber \\
m_{22}^2 = -(\beta_2v_2^2+\beta_3v^2)~;~&&\lambda_2=2(\beta_2+\beta_3)~; \nonumber \\
m_{12}^2=\frac{\beta_5}{2}v_1v_2~;~&& \lambda_3=(2\beta_3+\beta_4)~; \nonumber \\
\lambda_4=\frac{\beta_5+\beta_6}{2}-\beta_4~;~&&  \lambda_5=\frac{\beta_5-\beta_6}{2}~.
\label{connections}
\end{eqnarray}
At this point, it is interesting to note that in the parametrization of \Eqn{potential_notation1}, the combination
$m_{12}^2/(\sin\beta\cos\beta)$  instead of $m_{12}^2$ itself controls the nonstandard masses\cite{Gunion:2002zf}.
Thus the relation between $m_{12}^2$ and $\beta_5$ in \Eqn{connections} suggests that $\beta_5$ is a better 
parameter for tracking the effect of soft breaking. In passing, we note that in the limit $\lambda_5=0$ (in \Eqn{potential_notation1}) or equivalently $\beta_5=\beta_6$ (in \Eqn{potential_notation2}),  the symmetry of
the 2HDM potential is enhanced from softly broken $Z_2$ to  softly broken $U(1)$ under which $\phi_1 \to \phi_1$
and $\phi_2 \to e^{i\theta}\phi_2$. This $U(1)$ symmetry will be relevant in our future discussions.

To get the mass eigenstates, we expand the scalar doublets around their vevs as follows:
\begin{eqnarray}
\phi_i = \left(\begin{array}{c}
\chi_i^+ \\
\frac{1}{\sqrt{2}} (v_i + h_i + i\eta_i)
\end{array} \right) \,.
\label{field_expansion}
\end{eqnarray}
Our assumption that CP is a good symmetry of the scalar potential actually allows us
to define neutral scalar eigenstates that are also eigenstates of CP.
In total, there are five physical eigenstates: two CP-even scalars $(h,H)$,
one CP-odd scalar $(A)$ and a pair of charged scalars $(H^\pm)$
along with three Goldstones $(G^\pm,G^0)$ which will be absorbed to give masses to the
 SM gauge bosons $(W^\pm, Z)$. The rotations that lead us to the mass eigenstates are given below:
\begin{subequations}
\begin{eqnarray}
\left(\begin{array}{c}
G^\pm \\
H^\pm \end{array} \right) &=& 
\left(\begin{array}{cc}
c_\beta & s_\beta  \\
-s_\beta & c_\beta \end{array} \right) \left(\begin{array}{c}
\chi_1^\pm \\
\chi_2^\pm \end{array} \right) \,, \\
\left(\begin{array}{c}
G^0 \\
A \end{array} \right) &=& \left(\begin{array}{cc}
c_\beta & s_\beta  \\
-s_\beta & c_\beta \end{array} \right)\left(\begin{array}{c}
\eta_1 \\
\eta_2 \end{array} \right) \,, \\
\left(\begin{array}{c}
h \\
H \end{array} \right) &=& \left(\begin{array}{cc}
c_\alpha & s_\alpha  \\
-s_\alpha & c_\alpha \end{array} \right)\left(\begin{array}{c}
h_1 \\
h_2 \end{array} \right) \,.
\end{eqnarray}
\label{rotation}
\end{subequations}
where, $c_{\beta(\alpha)} \equiv \cos\beta(\alpha)$ and $s_{\beta(\alpha)} \equiv \sin\beta(\alpha)$.
The mixing angle of the CP-even sector is defined through the following relation:
\begin{eqnarray}
\tan 2\alpha = \frac{2(\beta_3+\frac{\beta_5}{4} )v_1v_2}
{\beta_1v_1^2 - \beta_2v_2^2 +(\beta_3 + \frac{\beta_5}{4})(v_1^2 - v_2^2)} \,.
\label{angle_alpha}
\end{eqnarray}

It is now instructive to count the number of free parameters in the scalar potential. Note that,
\Eqs{potential_notation1}{potential_notation2} both contain eight free parameters. In the notation
of \Eqn{potential_notation2}, these are $v_1$, $v_2$ and six $\beta_i$ couplings. We can trade $v_1$ and $v_2$
for $v = \sqrt{v_1^2 + v_2^2}$ and $\tan\beta$. Except for $\beta_5$, all other $\beta$ parameters may be traded
for four physical scalar masses ($m_h,m_H,m_A$ and $m_{H^+}$) and the angle, $\alpha$. The equivalence 
of these two sets of parameters is demonstrated by the following relations:
\begin{subequations}
\begin{eqnarray}
\beta_1 &=& \frac{1}{2v^2c_\beta^2}\left[m_H^2c_\alpha^2
  +m_h^2s_\alpha^2
  -\frac{s_\alpha c_\alpha}{\tan\beta}\left(m_H^2-m_h^2\right)\right]
-\frac{\beta_5}{4}\left(\tan^2\beta-1\right) \,, \\
\beta_2 &=& \frac{1}{2v^2s_\beta^2}\left[m_h^2c_\alpha^2
  +m_H^2s_\alpha^2
  -s_\alpha c_\alpha\tan\beta\left(m_H^2-m_h^2\right) \right]
-\frac{\beta_5}{4}\left(\cot^2\beta-1\right) \,, \\
\beta_3 &=& \frac{1}{2v^2}
\frac{s_\alpha c_\alpha}{s_\beta c_\beta}
\left(m_H^2-m_h^2\right) -\frac{\beta_5}{4} \,, \\
\beta_4 &=& \frac{2}{v^2} m_{H^+}^2 \,, \\
\beta_6 &=& \frac{2}{v^2} m_A^2 \,. \label{mA}
\end{eqnarray}
\label{mass_coup}
\end{subequations}
Among the eight redefined parameters that appear on the RHS of \Eqn{mass_coup}, not all are unknown. We 
already know $v=$ 246~GeV and under the assumption that the lightest CP-even Higgs is what has been found 
at the LHC, $m_h\approx$ 125~GeV is also known.

As a next level of simplification, we recall that the experimental values of the Higgs signal strengths into
different decay modes are showing increasing affinity towards the SM expectations \cite{ATLAS-conf-2014-009,CMS:2014ega}. This encourages us to work in the {\em alignment limit} \cite{Gunion:2002zf}:
\begin{eqnarray}
\beta - \alpha = \frac{\pi}{2} \,,
\label{alignment}
\end{eqnarray} 
which means, $h$ will have the exact SM-like tree-level couplings with the fermions and the vector bosons.
The recent global fits of the LHC data in the 2HDM context \cite{Coleppa:2013dya,Chen:2013rba,Craig:2013hca, Eberhardt:2013uba,Dumont:2014wha,Dumont:2014kna,Bernon:2014vta}
certifies that \Eqn{alignment} is indeed a reasonable assumption. In what follows, we will work
exclusively in the alignment limit. Therefore, in this limit, we are left with five unknown parameters 
($m_H,m_A$, $m_{H^+}$, $\beta_5$ and $\tan\beta$) which will be constrained from the requirement
of high scale stability of the 2HDM potential. We will see that this requirement in association with
the constraints from electroweak $T$-parameter entail strong correlations between most of the remaining
parameters making the 2HDM more constrained than ever.

\subsection{Theoretical constraints from vacuum stability and unitarity}
First we have to ensure that there is not any direction in the field space along which the potential
becomes infinitely negative. In the parametrization of \Eqn{potential_notation1}, the necessary
and sufficient conditions for the potential to be bounded from below read \cite{Deshpande:1977rw,Klimenko:1984qx,Kastening:1992by, Maniatis:2006fs}
\begin{subequations}
\begin{eqnarray}
{\rm VSC1:} && \lambda_1 > 0, \label{sta1} \\
{\rm VSC2:} && \lambda_2 > 0, \label{sta2}\\
{\rm VSC3:} && \lambda_3 + \sqrt{\lambda_1\lambda_2} > 0, \label{sta3}\\
{\rm VSC4:} && \lambda_3 + \lambda_4 - |\lambda_5| + \sqrt{\lambda_1\lambda_2} > 0. \label{sta4}
\end{eqnarray}
\label{vsc}
\end{subequations}
The $S$-matrix eigenvalues that will be constrained from unitarity of the scattering amplitudes are also
listed below \cite{Maalampi:1991fb, Kanemura:1993hm, Akeroyd:2000wc, Horejsi:2005da}:
\begin{subequations}
\begin{eqnarray}
a_1^\pm &=& \frac{3}{2}(\lambda_1 + \lambda_2) \pm \sqrt{\frac{9}{4} (\lambda_1 - \lambda_2)^2 + \left(2\lambda_3 + \lambda_4 \right)^2} \,, \label{un1}\\
a_2^\pm &=& \frac{1}{2}(\lambda_1 + \lambda_2) \pm \frac{1}{2}\sqrt{(\lambda_1 - \lambda_2)^2 + 4\lambda_4^2} \,, \label{un2}\\
a_3^\pm &=& \frac{1}{2}(\lambda_1 + \lambda_2) \pm \frac{1}{2}\sqrt{(\lambda_1 - \lambda_2)^2 + 4\lambda_5^2} \,, \label{un3}\\
b_1^\pm &=& \lambda_3 + 2\lambda_4 \pm 3\lambda_5 \,, \label{un4}\\
b_2^\pm &=& \lambda_3 \pm \lambda_5 \,, \label{un5}\\
b_3^\pm &=& \lambda_3 \pm \lambda_4 \,. \label{un8}
\end{eqnarray}
\label{unitarity}
\end{subequations} 
The requirement of tree unitarity then restricts the above eigenvalues as follows:
\begin{eqnarray}
|a_i^\pm|\,,~ |b_i^\pm| \leq 16\pi \,.
\label{uni}
\end{eqnarray}
We impose that the inequalities (\ref{vsc}) and (\ref{uni}) should be satisfied at all energies
between the electroweak and Planck scales. The renormalization group (RG) equations that we
use to calculate the lambdas at any intermediate energy scale are given in Appendix~\ref{RGE_all}.

Additionally it is also important to check whether the minimum defined by $v_1$ and $v_2$ is a
global minimum or not. The condition for the global minimum, in the notation of \Eqn{potential_notation1},
is given by\cite{Barroso:2012mj,Barroso:2013awa},
\begin{eqnarray}
{\mathscr D}= m_{12}^2\left(m_{11}^2-m_{22}^2\sqrt{\frac{\lambda_1}{\lambda_2}}\right) \left(\tan\beta -\sqrt[4]{\frac{\lambda_1}{\lambda_2}}\right) > 0 \,.
\label{global_m}
\end{eqnarray}

\subsection{Experimental constraints}
In addition to the theoretical constraints mentioned above, we also take the following
experimental facts into account.
\begin{itemize}
\item There is a very strong lower limit on $m_{H^+}$, for Type~II models, arising mainly due to
the excellent agreement of the $b\to s\gamma$ branching ratio with the SM prediction. Because of this
we take $m_{H^+}>300$ GeV \cite{Mahmoudi:2009zx,Deschamps:2009rh} for Type~II models. However, if $\tan\beta>1$, there is no such bound for Type~I models from flavor data \cite{Mahmoudi:2009zx}. 
Therefore, for Type~I models, we only consider the direct search limit $m_{H^+}>80$ GeV \cite{Searches:2001ac}.
 At this stage, it should be noted that since the Yukawa couplings in the
quark sector are the same for Type~II and Y models and the top-Yukawa gives the dominant fermionic
contribution in the RG equations, whatever we comment on Type~II models in this paper will
eventually hold for Type~Y models too. The same is true for Type~I and X models.
\item The oblique  $T$-parameter can restrict the splitting between 
the heavy scalar masses. In the 2HDM alignment limit, the expression for the new physics contribution to the $T$-parameter can be expressed as \cite{He:2001tp,Grimus:2007if}
\begin{eqnarray}
\Delta T &=&  \frac{1}{16\pi\sin^2\theta_wM_W^2} \left[F(m_{H^+}^2,m_H^2) + F(m_{H^+}^2,m_A^2) - F(m_H^2,m_A^2) \right] \,, \\
{\rm with,}~~~~
F(x,y)&=&\left\{
\begin{array}{cc}
\frac{x+y}{2} - \frac{xy}{x-y}\ln\left(\frac{x}{y} \right) & {\rm for}~~ x\neq y \,, \\
0 & ~{\rm for}~~x=y \,.
\end{array}
\right.
 \label{tparam}
\end{eqnarray}
Taking the new physics contribution to the $T$-parameter as \cite{Baak:2013ppa}
\begin{eqnarray}
\Delta T = 0.05 \pm 0.12 \,,
\end{eqnarray}
we use the $2\sigma$ uncertainty range around the mean value for our numerical constraints.
Note that the function $F(x,y)$ is symmetric under $x \leftrightarrow y$ and is sensitive only to the difference $|x-y|$.
Thus $\Delta T=0$ when either $m_{H^+}=m_H$ or $m_{H^+}=m_A$. In these cases $\Delta T$ puts no
constraints on the heavy scalar masses. But if, for some reason, $m_H\approx m_A$ then $\Delta T$ severely
restricts the splitting between charged and neutral scalar masses.
\end{itemize}

\begin{figure}[h]
	\centering
	\begin{tabular}{c c}
		\includegraphics[width=7cm,height=5cm]{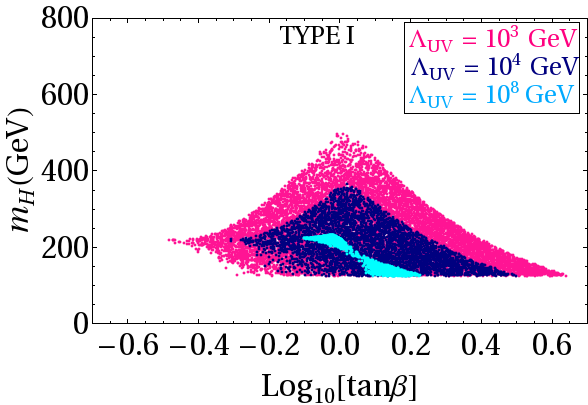} & 
		\includegraphics[width=7cm,height=5cm]{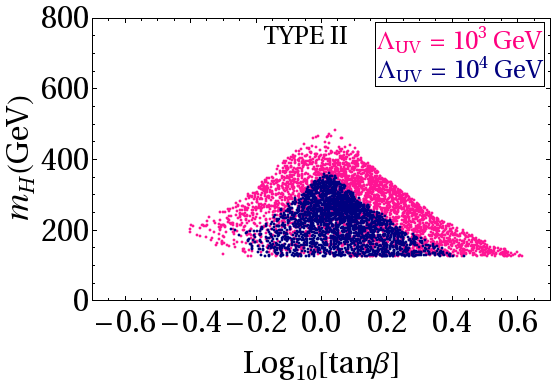} \\
    	\includegraphics[width=7cm,height=5cm]{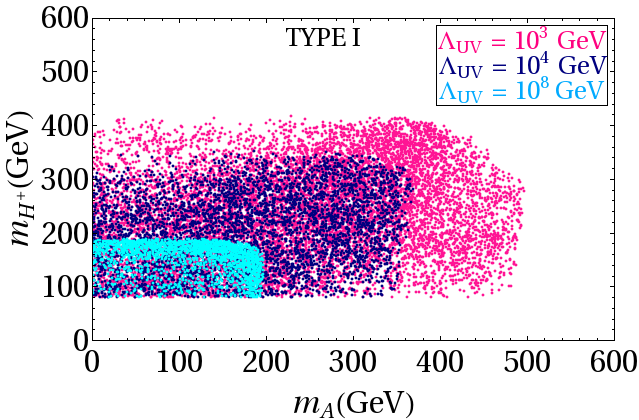} & 
    	\includegraphics[width=7cm,height=5cm]{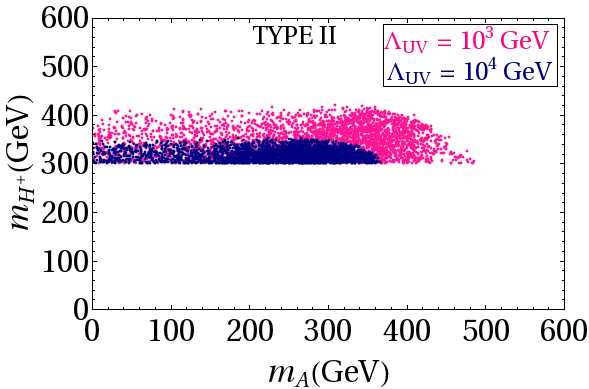} 
   	\end{tabular}
	\caption{\em (Exact $Z_2$) Allowed points in the ${\rm Log}_{10}(\tan\beta)$-$m_H$ and $m_A$-$m_{H^+}$
	planes for the 2HDM potential to be stable up to $\Lambda_{\rm UV}$. The points in different colors
	correspond to different choices for $\Lambda_{\rm UV}$ which appear in the legends.}
	\label{exact_z2}
\end{figure}
\section{Numerical analysis and results}\label{results}
For the purpose of analysis, we choose the set $\{v, \tan\beta, \alpha, m_h, m_H, m_A, m_{H^+}, \beta_5\}$ which appears on the RHS of \Eqn{mass_coup} to be
our independent set of eight parameters to describe the 2HDM scalar potential. Among them, we set $v=246$ GeV and $\alpha =\beta-\pi/2$. In the case of an exact $Z_2$ symmetry we also set $\beta_ 5=0$ and perform a
random scan for the rest of the parameters in the following ranges:
\begin{eqnarray}
&& \tan\beta \in [0.1, 10] ~,~ m_h \in [124, 126]~{\rm GeV} ~,~ m_H \in[m_h, 1500]~{\rm GeV} ~,~ 
m_A \in [0, 1500]~{\rm GeV} ~, \nonumber \\
&& m_{H^+} \in [80, 1500]~{\rm GeV} ~~ {\rm (for~ Type~ I~ models)} \,, ~~
m_{H^+} \in [300, 1500]~{\rm GeV}~~ {\rm (for~ Type~ II~ models)} \,.  
\label{scanrange}
\end{eqnarray}
When the $Z_2$ symmetry is softly broken, we find it convenient to introduce the mass parameter
\begin{eqnarray}
m_0 = \sqrt{\frac{1}{2}\beta_5v^2} \,,
\label{m0}
\end{eqnarray}
and vary $m_0$ in the range [0, 1500] GeV. In the case of a softly broken $U(1)$ symmetry, we impose
$m_0=m_A$. 

Next we use \Eqn{mass_coup} to convert our set of parameters into the set of $\beta_i$s followed by \Eqn{connections} to
convert the $\beta_i$s again into $\lambda_i$s. We then use the RG equations for $\lambda_i$s given in Appendix \ref{RGE_all} to compute them at any intermediate energy scale and check that the unitarity and stability conditions hold all the way up to
the Planck scale. We set the top quark pole mass at 173.3~GeV \cite{ATLAS:2014wva} for our numerical analysis. Our observations for different cases appear in the following subsections.

\subsection{Exact \texorpdfstring{$Z_2$}{TEXT}} 
In this case setting $\beta_5=0$, we scan the rest of the parameters in the range specified previously. The result
has been displayed in Fig.~\ref{exact_z2}. A couple of noteworthy features that emerge from this figure are
given below:
\begin{enumerate}[($i$)]
\item The value of $\tan\beta$ is bounded and the bound depends on the energy scale, $\Lambda_{\rm UV}$, up
to which stability is demanded. As we increase $\Lambda_{\rm UV}$, the allowed parameter space shrinks
continuously. It should be noted that a limit on $\tan\beta$ for exact $Z_2$ symmetry is not surprising at
all and is a direct consequence of unitarity and stability at the electroweak scale \cite{Swiezewska:2012ej,Das:2015qva}.
\item It is found that, for the non-inert cases, neither Type~I nor Type~II models can be absolutely stable all the way up to
the Planck scale when the $Z_2$ symmetry is exact in the potential. This result is in agreement with the previous analysis of Ref.~\cite{Chakrabarty:2014aya} in the context of exact $Z_2$ symmetry. We have checked for each order of magnitude
in $\Lambda_{\rm UV}$ and noticed that Type~I models can remain stable up to a maximum of $10^8$~GeV whereas Type II models can be stable only up to $10^4$~GeV. This difference between Type~I and Type~II
models may be understood by noting that stability up to $10^8$~GeV for Type~I models requires a light
charged scalar, $m_{H^+}\sim 180$ GeV (see lower left panel of Fig.~\ref{exact_z2}), which is not allowed
for Type~II models from $b\to s\gamma$.
\end{enumerate}
Now that we know 2HDMs with exact $Z_2$ symmetry fail to maintain stability up to the Planck scale, it
is time to investigate whether the introduction of a soft symmetry breaking parameter can improve the situation.

\begin{figure}[htbp!]
     	\centering
     	\begin{tabular}{c c}
     		\includegraphics[width=7cm,height=5cm]{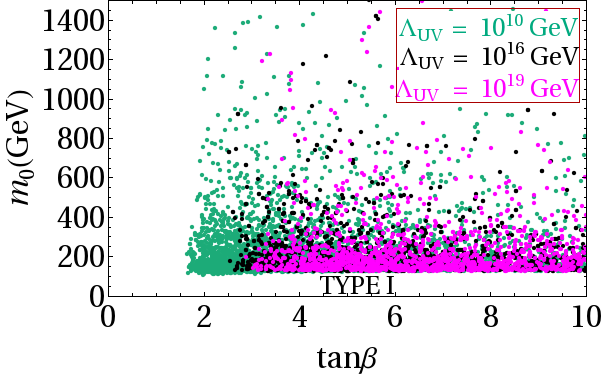} & 
     		\includegraphics[width=7cm,height=5cm]{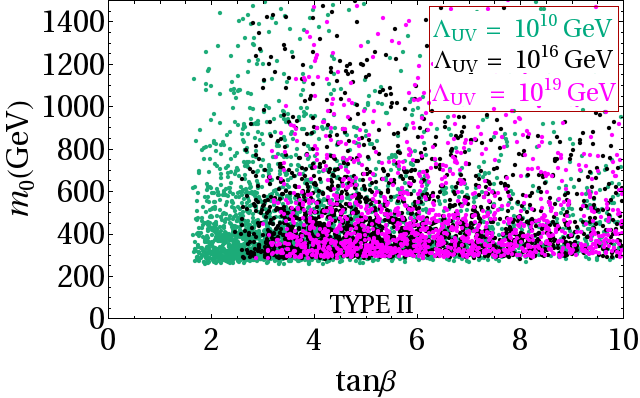} 
     	\end{tabular}
     	\caption{\em (Softly broken $Z_2$) Allowed points in $\tan\beta$-$m_0$ plane for the 2HDM potential to be stable up to $\luv$. The points in different colors	correspond to different choices for $\Lambda_{\rm UV}$ which appear in the legends.}
     	\label{broken_z2}
\end{figure}
\subsection{Softly broken \texorpdfstring{$Z_2$}{TEXT}}
Here we study the effects of nonzero $\beta_5$. For convenience we have encoded the information of $\beta_5$ into
the mass parameter, $m_0$, through \Eqn{m0}. The same analysis as in the case of exact $Z_2$ symmetry is then
performed for three representative choices of $\Lambda_{\rm UV}$, namely, $10^{10}$, $10^{16}$ and $10^{19}$~GeV.
We exhibit our results in Figs.~\ref{broken_z2} and \ref{broken_z2_mass}. Some intriguing features that emerge from these figures are listed below.
\begin{enumerate}[$(a)$]
\item From Figs.~\ref{broken_z2} and \ref{broken_z2_mass}, we note that it is indeed possible for a 2HDM potential
to remain stable all the way up to the Planck scale in the presence of a soft symmetry breaking parameter. But certain
conditions need to be satisfied for this to happen. These conditions which we describe below, together define the 
{\em stable alignment limit} for 2HDMs.
\begin{figure}[h!]
     	\centering
     	\begin{tabular}{c c}
     		\includegraphics[width=7cm,height=5cm]{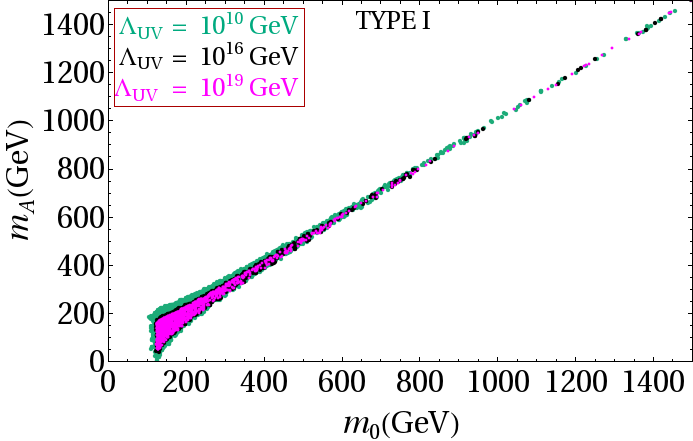} & 
     		\includegraphics[width=7cm,height=5cm]{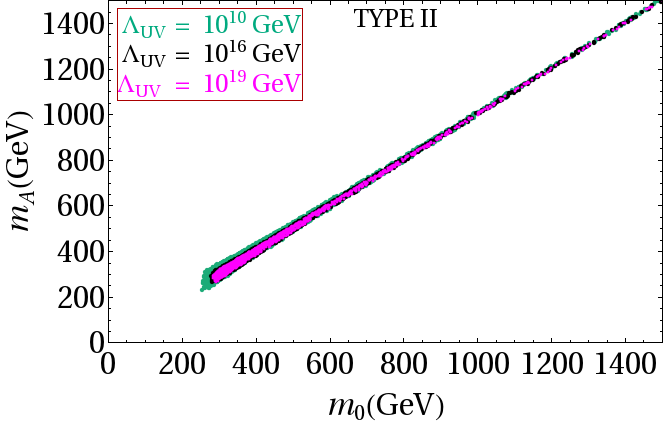} \\
     		\includegraphics[width=7cm,height=5cm]{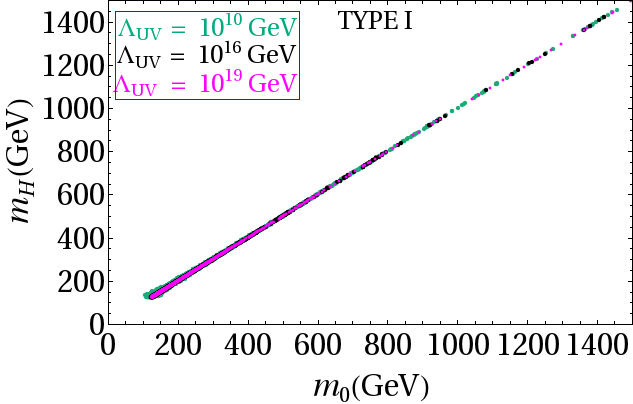} & 
     		\includegraphics[width=7cm,height=5cm]{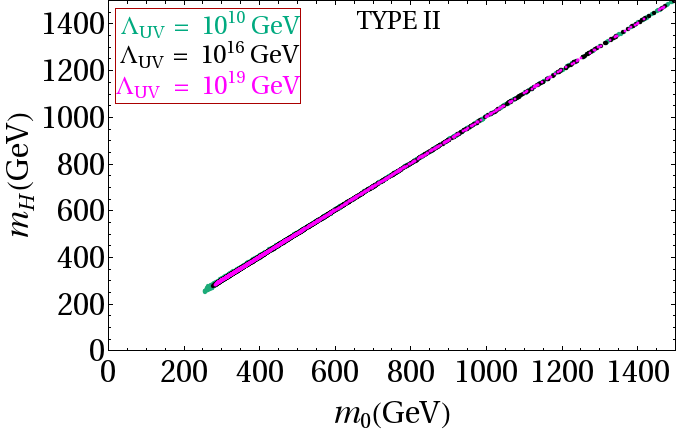} \\
     		\includegraphics[width=7cm,height=5cm]{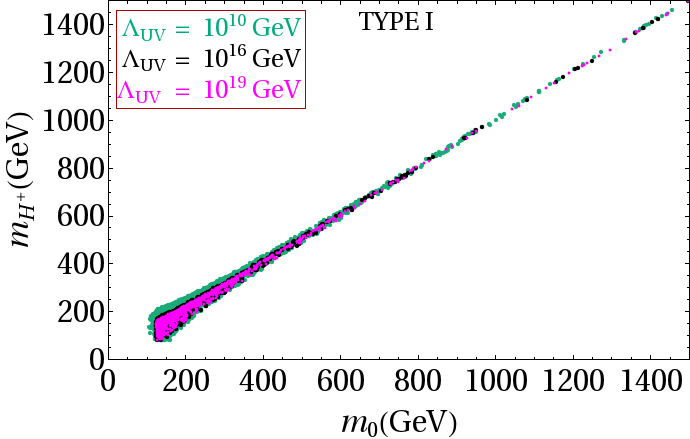} & 
     		\includegraphics[width=7cm,height=5cm]{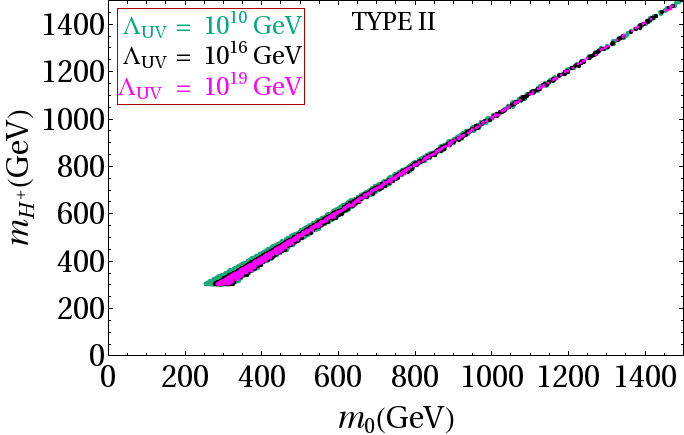}
     	\end{tabular}
     	\caption{\em (Softly broken $Z_2$) Allowed points in $m_0$-$m_H$, $m_0$-$m_{H^+}$, $m_0$-$m_A$
     		 planes for the 2HDM potential to be stable up to $\luv$. The points in different colors	correspond to different choices for $\Lambda_{\rm UV}$ which appear in the legends. }
     	\label{broken_z2_mass}
\end{figure}
\item From Fig.~\ref{broken_z2} we see that for the 2HDM potential 
to be stable up to $\Lambda_{\rm UV}$, there must
exist a lower bound on $\tb$. For example, when $\luv=10^{19}$~GeV 
this bound reads $\tb \gtrsim 3$. We also note
from Fig.~\ref{broken_z2} that there is a lower limit on $m_0$
 (or equivalently $\beta_5$) too. For Type~I models we read
$m_0\gtrsim 120$~GeV whereas for Type~II models we obtain 
$m_0\gtrsim 280$~GeV. For the energy scales that we are
concerned with, we observe that these limits on $m_0$ for 
Type~I and II models are essentially independent of $\luv$.
We emphasize that this is one of the important differences between our result
and that of Ref.~\cite{Chakrabarty:2014aya} where the stability of 2HDMs with
softly broken $Z_2$ symmetry was analyzed including $\tb=2$ without realizing that such a low
value of $\tb$ is forbidden when the alignment limit is exact.
\item From Fig.~\ref{broken_z2_mass} we notice that the requirement of absolute stability up to $\luv$ entails
a strong correlation between $m_0$ and the other nonstandard masses. In fact, a closer scrutiny of the plots reveals that all the nonstandard scalar masses have to be nearly degenerate with $m_0$.
\item Observations $(b)$ and $(c)$ together lead us to the
 conclusion that, in the {\em stable alignment limit}, a 2HDM
becomes completely determined by only two nonstandard
 parameters, namely, $\tb$ and a mass parameter ($m_0$).
Additionally, both of them are bounded from below:
\begin{eqnarray}
\tb\gtrsim 3 \,, ~~ m_0\gtrsim 120~{\rm GeV}~({\rm Type~I})\,, ~~ m_0\gtrsim 280~{\rm GeV}~({\rm Type~II})\,.
\end{eqnarray}
\item We have also checked that above conclusions do not crucially
 depend on the input parameters, especially the top-quark
mass. To be quantitative, instead of the central value of 
$173.3$~GeV, if we consider the $2\sigma$ lower limit of the
top-quark pole mass, 171.8~GeV \cite{ATLAS:2014wva}, the lower limit on $\tb$ changes to $\tb\gtrsim 2.8$.
\end{enumerate}

We feel that an {\it a posteriori} explanation of most of the above features is possible, at least on a qualitative level.
In the following we present, one by one, the steps of our argument that helps us apprehend different
characteristics of Figs.~\ref{broken_z2} and \ref{broken_z2_mass}.
\begin{enumerate}[$(1)$]
\item First we note from \Eqn{lambda_rg} in association with Eqs.~(\ref{l5_rg_l}), (\ref{l5_rg_g}), (\ref{l5_yk_t1})
and (\ref{l5_yk_t2}) that the evolution of $\lambda_5$ is proportional to itself. This is not surprising because,
as mentioned earlier,
in the absence of $\lambda_5$ the symmetry of the scalar potential is enhanced to a global $U(1)$. Thus if we
start with $\lambda_5=0$ at the electroweak scale, it will remain zero at all energy scales. But any initial nonzero 
value of $\lambda_5$ will cause it to grow with energy. This may eventually jeopardize the
vacuum stability condition~(\ref{sta4}) at high energy scales. Therefore, it will be no wonder if, by demanding
high scale stability of the 2HDM potential, we are led to the $U(1)$ limit, $\lambda_5\approx 0$. Evidently, this
limit will have the following implication on the masses (using \Eqn{mA} and remembering that  
$\lambda_5\approx 0$ implies $\beta_5\approx \beta_6$):
\begin{eqnarray}
m_A^2 \approx m_0^2 \,,
\end{eqnarray}
which has been clearly depicted by the first row of plots in Fig.~\ref{broken_z2_mass}.
\item Next important thing to note is that unitarity and stability conditions at the electroweak scale
imply, among other things, \cite{Bhattacharyya:2013rya,Das:2015qva}
\begin{eqnarray}
0<(m_H^2-m_0^2)(\tan^2\beta+\cot^2\beta)+2m_h^2<\frac{32\pi v^2}{3} \,.
\label{ineq}
\end{eqnarray}
This inequality has been plotted in Fig.~\ref{inequality} for three different values of $\tb$. It is obvious
that for $\tb$ away from unity, the inequality~(\ref{ineq}) renders a degeneracy between $m_H$ and
$m_0$. This explains the second row of Fig.~\ref{broken_z2_mass}.
\begin{figure}[htbp!]
\begin{minipage}{0.46\textwidth}
\centerline{\includegraphics[width=8cm,height=6cm]{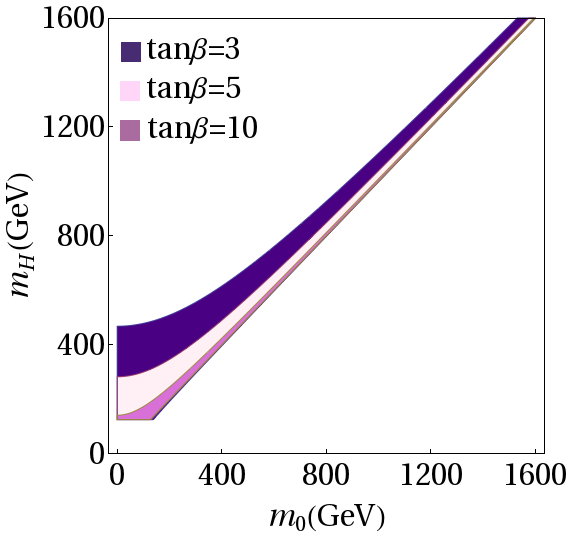}}
\caption{\em Allowed regions in the $m_0$-$m_H$ plane from the inequality (\ref{ineq}) 
for three different values of $\tan\beta$.}
\label{inequality}
\end{minipage}
\hfill
\begin{minipage}{0.46\textwidth}
\centerline{\includegraphics[width=8cm,height=6cm]{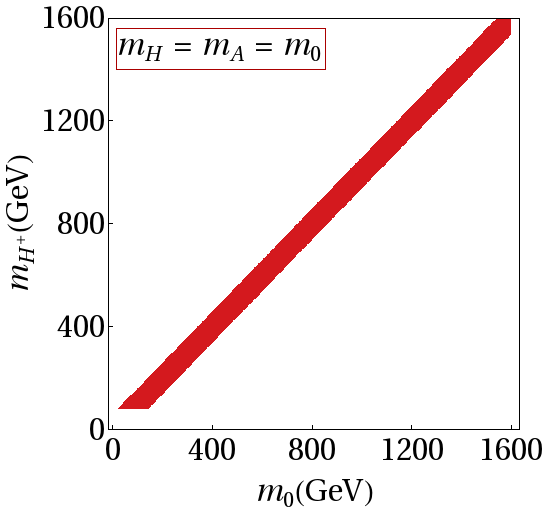}}
\caption{\em Allowed region at 95\% C.L. from the experimental limit on $\Delta T$ under the
assumption $m_H=m_A=m_0$.}
\label{t-parameter}
\end{minipage}
\end{figure} 
\item Collecting $(1)$ and $(2)$ together we get $m_H^2\approx m_A^2\approx m_0^2$. Feeding this
information to \Eqn{tparam}, we obtain the following expression for $\Delta T$:
\begin{eqnarray}
\Delta T  = \frac{1}{8\pi\sin^2\theta_wM_W^2}F\left(m_{H^+}^2, m_0^2\right) \,.
\label{del-t}
\end{eqnarray}
Since $F\left(m_{H^+}^2, m_0^2\right)$ restricts the splitting $|m_{H^+}^2-m_0^2|$, the experimental
limit on $\Delta T$ will impart the degeneracy between $m_{H^+}$ and $m_0$ as shown in 
Fig.~\ref{t-parameter}. This explains the third row of Fig.~\ref{broken_z2_mass}.

Thus, up to this point, we have
\begin{eqnarray}
m_0^2 \approx m_A^2 \approx m_H^2 \approx m_{H^+}^2 \,,
\label{degeneracy}
\end{eqnarray}
which summarizes the essential features of Fig.~\ref{broken_z2_mass} that the 2HDM can be
described by a single nonstandard mass parameter. \Eqn{degeneracy} also explains
the lower bounds on $m_0$ in Fig.~\ref{broken_z2}. Since $m_{H^+}>300$~GeV for Type~II
2HDMs, the degeneracy of \Eqn{degeneracy} destroys the possibility of having a light pseudoscalar
for Type~II models. This result is apparent from the upper right panel of Fig.~\ref{broken_z2_mass}.
\begin{figure}[htbp!]
\begin{minipage}{0.46\textwidth}
\centerline{\includegraphics[width=7.5cm,height=6cm]{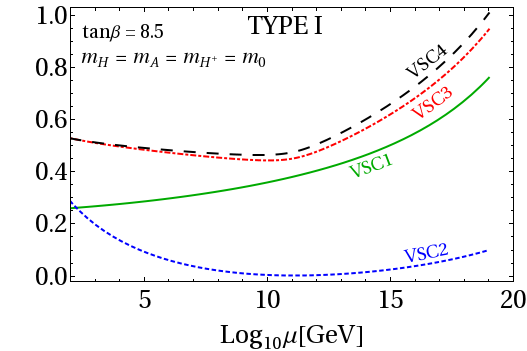}}
\caption{\em  (Softly broken $Z_2$) Running of stability conditions of Eq.~(\ref{vsc}) 
in the exactly degenerate scenario assuming $m_h=126$~GeV. In this scenario, these runnings are independent
of $m_0$.}
\label{sta_run}
\end{minipage}
\hfill
\begin{minipage}{0.46\textwidth}
\centerline{\includegraphics[width=8cm,height=6cm]{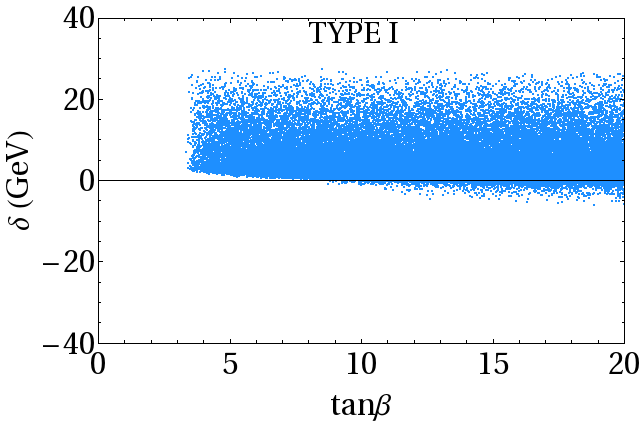}}
\caption{\em (Softly broken $Z_2$) Allowed points in $\tan\beta$-$\delta$ plane 
 for $\luv=10^{19}~\rm GeV$ in a Type~I 2HDM with $m_H=m_A=m_0$, and
 $ m_{H^+} = m_0+\delta$.}
\label{tb_delta}
\end{minipage}
\end{figure} 
\item To understand the lower limit on $\tb$ we turn our attention to the RG evolution equation of $\lambda_2$
which is given by \Eqn{lambda_rg} in association with Eqs.~(\ref{l2_rg_l}), (\ref{l2_rg_g}), (\ref{l2_yk_t1})
and (\ref{l2_yk_t2}). Since we have assumed $\phi_2$ to be the doublet that gives masses to the up-type 
quarks, $\lambda_2$ will face the negative pull of the top-Yukawa ($h_t$). Note that in the limit of exact degeneracy,
$m_0^2 = m_A^2 = m_H^2 = m_{H^+}^2$, we have at the electroweak scale, (using \Eqn{mass_coup} 
followed by \Eqn{connections} in the alignment limit)
\begin{eqnarray}
\lambda_1 =\lambda_2= \lambda_3 = \frac{m_h^2}{v^2} \,, ~~~{\rm and},~~ \lambda_4=\lambda_5=0 \,.
\end{eqnarray}
Using these, we may simplify the RG equation for $\lambda_2$, at the electroweak scale, as follows:
\begin{eqnarray}
{\cal D}\lambda_2 &=& 16\lambda_2^2 -3\left(3g^2+{g^\prime}^2\right)\lambda_2 + \frac{3}{4}\left(3g^4 + {g^\prime}^4 + 2g^2{g^\prime}^2\right) 
+12h_t^2\lambda_2 - 12h_t^4\,.
\label{l2_RG}
\end{eqnarray}
Notice the striking similarity between the above equation and the SM running of $\lambda$ that appears
in \Eqn{SM_RG}. One difference is that ${\cal D}\lambda_2$ receives additional contributions on the positive
side (compare $16\lambda_2^2$ with $12\lambda^2$ in the SM case) due to the presence of extra quartic
couplings. On the contrary, $h_t$ in the 2HDM case is $\sim \sqrt{2}m_t/(v\sin\beta)$ which is
larger than the SM value of $h_t\sim \sqrt{2}m_t/v$. Therefore, compared to the SM case the 
negative drag of $h_t$ is enhanced in 2HDMs. This effect drives $\lambda_2$ to negative values
violating vacuum stability condition (\ref{sta2}) at high energies unless $\sin\beta$ is large enough
to dilute the effect of the term, $- 12h_t^4$, sufficiently. This is the origin of the lower limit on $\tb$.
Choosing $\tb=8.5$ for illustration, we have displayed this effect in Fig.~\ref{sta_run}. 
There we see how the evolution of the dotted (blue) line marked as VSC2 (which is nothing but $\lambda_2$) marginally survives becoming negative at an intermediate energy scale. Take $\tb$ too low and this line
goes below zero spoiling the stability condition, $\lambda_2>0$. It is worth noting that when all the nonstandard
masses are exactly degenerate, evolutions of the $\lambda_i$s depend only on $\tb$ and not on the mass
parameter, $m_0$.
\begin{figure}[ht!]
	\centering
\includegraphics[scale=0.75]{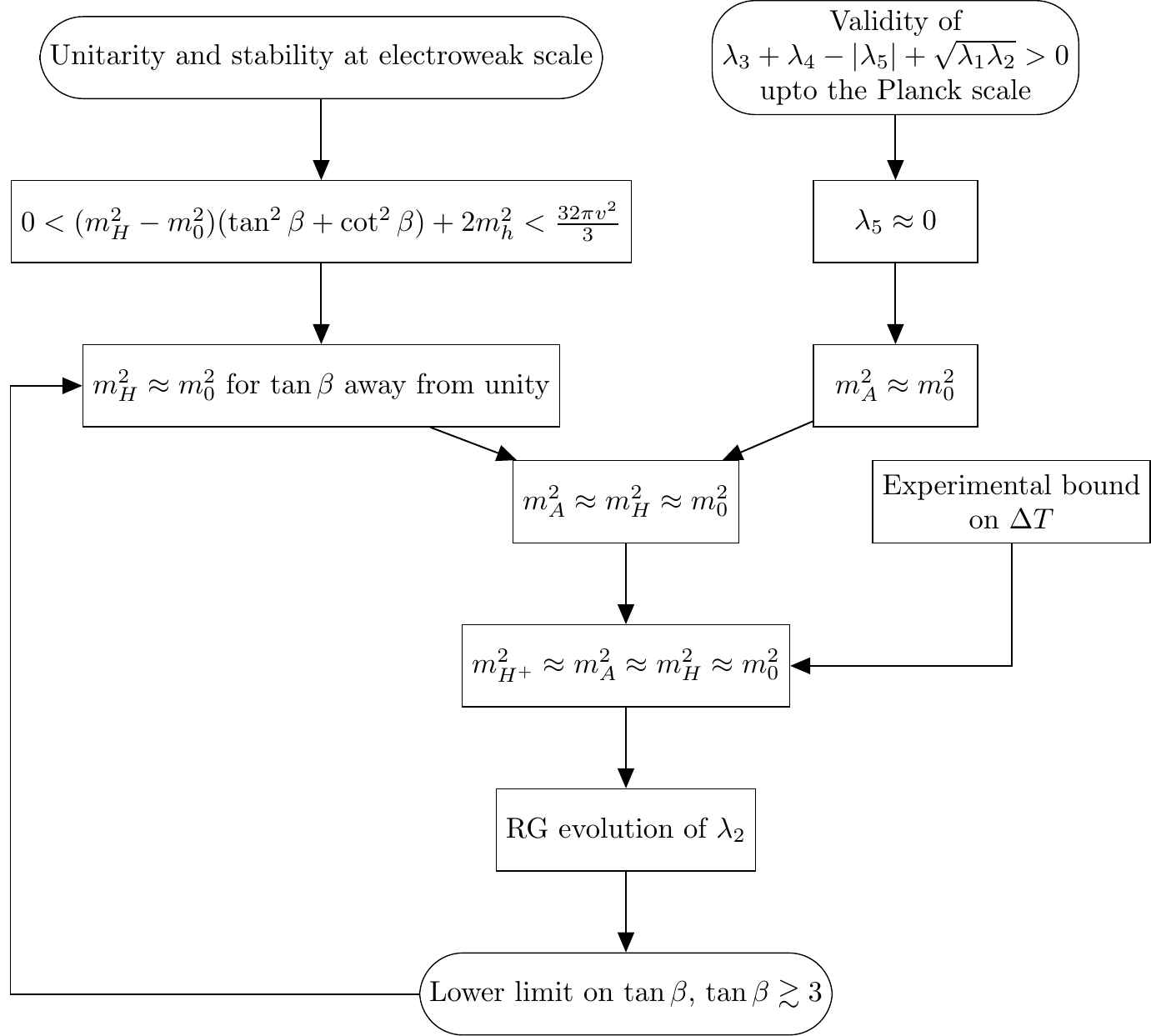}
\caption{\em Chain of arguments that helps us make sense of the main features of Figs.~\ref{broken_z2}
	and \ref{broken_z2_mass}.}
	\label{flowchart}
	\end{figure}
\item With exact degeneracy, the lower bound on $\tb$ turns out to be $\tb\gtrsim 8$ for $\luv=10^{19}$~GeV.
 But from Fig.~\ref{broken_z2} we see that even lower values of $\tb$ are allowed. To understand this, we need
to investigate the evolution of $\lambda_2$ in greater detail. Instead of exact degeneracy, we now consider
the following limit:
\begin{eqnarray}
m_0 = m_A = m_H \,, ~~~ {\rm and,} ~~ m_{H^+}^2 = m_0^2 + \Delta \,.
\end{eqnarray}
In this limit,
\begin{eqnarray}
\lambda_1 = \lambda_2 =\frac{m_h^2}{v^2} \,, ~~ \lambda_3=\lambda_2+\frac{2\Delta}{v^2} \,, ~~
\lambda_4 = -\frac{2\Delta}{v^2} \,, ~~ \lambda_5 = 0 \,.
\end{eqnarray}
Using these, we can rearrange the terms that appear on the RHS in the RG equation for $\lambda_2$, to obtain
\begin{eqnarray}
{\cal D}\lambda_2 &=& 14\lambda_2^2 +2\left(\lambda_2+\frac{2\Delta}{v^2} \right)^2 -3\left(3g^2+{g^\prime}^2\right)\lambda_2 + \frac{3}{4}\left(3g^4 + {g^\prime}^4 + 2g^2{g^\prime}^2\right) 
+12h_t^2\lambda_2 - 12h_t^4\,.
\label{l2_RG1}
\end{eqnarray}
Comparing this with \Eqn{l2_RG} we see that a positive value of $\Delta$ aids in the positive terms
allowing lower values of $\tb$. But the fact that only very small values of $\Delta$ can be permitted
from unitarity \cite{Das:2015qva} and $T$-parameter puts a lower limit on $\tb$ anyway. In 
Fig.~\ref{tb_delta} we have illustrated how the difference $\delta\equiv m_{H^+}-m_0 \approx \Delta/(2m_0)$
depends on $\tb$. There we can see that to allow $\tb\lesssim 8$ we will need $\delta>0$, {\it i.e.},
$m_{H^+}> m_0$.
\end{enumerate}

We have summarized our arguments above compactly in the form of a flowchart in Fig.~\ref{flowchart}.
Before moving on, we intend to make some remarks on the structure of the potential in the {\em stable
alignment limit}. We note that in the limit, $m_0^2 = m_A^2 = m_H^2 = m_{H^+}^2$, the potential
of \Eqn{potential_notation1} takes the following simple form:\footnote{Similar potential can also be motivated from an $SO(5)$ symmetry \cite{Dev:2014yca}.}
\begin{eqnarray}
 V &=& m_{11}^2 \phi_1^\dagger\phi_1 +
m_{22}^2\phi_2^\dagger\phi_2 -m_{12}^2\left( \phi_1^\dagger\phi_2
+{\rm h.c.} \right) +\frac{\lambda_1}{2} \left(\phi_1^\dagger\phi_1+\phi_2^\dagger\phi_2 
\right)^2  \,.
\label{potential_sal}
\end{eqnarray}
This potential now contains four parameters among which two, disguised as $v$ and $m_h$, have
been measured. Two other parameters, in the form of $\tb$ and $m_0$, remain to be determined to
fix the model completely. Note that the symmetry in the quartic terms of \Eqn{potential_sal} is more
than a mere $U(1)$, in fact, the symmetry is enhanced to a global $U(2)$ under which 
$(\phi_1,~\phi_2)^T \to U(2)(\phi_1,~\phi_2)^T$. But this $U(2)$ symmetry is explicitly broken
in the Yukawa terms. Due to this, the structure of the potential in \Eqn{potential_sal} is not
stable under RG. Neither the correlation $\lambda_1=\lambda_2=\lambda_3$ nor
the equality $\lambda_4=0$ is maintained at higher energies. What remains valid at any scale is
the equality $\lambda_5=0$ because the $U(1)$ symmetry that prevails in the quartic part of the scalar potential
in the absence of $\lambda_5$ is also preserved in the Yukawa sector by construction. Therefore, we
conclude that when high scale stability of the 2HDM potential needs to be ensured, softly broken
$U(1)$ symmetry should be a more natural choice, to tackle tree-level FCNC, than the conventional
$Z_2$ symmetry.

\begin{subequations}
Now we want to check whether the condition (\ref{global_m}) for the global minimum holds in the stable alignment
limit or not. First we remember that, in this limit, $m_0$ is positive (see Fig.~\ref{broken_z2}) which implies
$m_{12}^2>0$. We have also found $\lambda_1\approx\lambda_2\approx\lambda_3$ with $\tb \gtrsim 3$. These
imply
\begin{eqnarray}
\left(\tan\beta -\sqrt[4]{\frac{\lambda_1}{\lambda_2}}\right) > 0 \,.
\end{eqnarray}
The remaining factor in \Eqn{global_m} can be evaluated, in the stable alignment limit, as follows:
\begin{eqnarray}
m_{11}^2-m_{22}^2\sqrt{\frac{\lambda_1}{\lambda_2}} &\approx & m_{11}^2-m_{22}^2 = \beta_2v_2^2-\beta_1v_1^2 ~~~~~{\rm [Using ~\Eqn{connections}]} \\
&=& m_A^2\left(\sin^2\beta-\cos^2\beta \right) ~~~~~~~~~~~~{\rm [Using ~\Eqn{mass_coup}~ with~ m_0\approx m_A]} \\
&=& m_A^2\cos^2\beta(\tan^2\beta-1) > 0 ~~~~~~~{\rm [\because \tan\beta\gtrsim 3]}
\end{eqnarray}
Therefore we conclude that, in the stable alignment limit, condition~(\ref{global_m}) is satisfied automatically and
hence, existence of the global minimum is guaranteed.
\end{subequations}

\subsection{Softly broken \texorpdfstring{$U(1)$}{TEXT}}
\begin{figure}[ht!]
	\centering
		\begin{tabular}{c c c}
		\includegraphics[width=5.2cm,height=4cm]{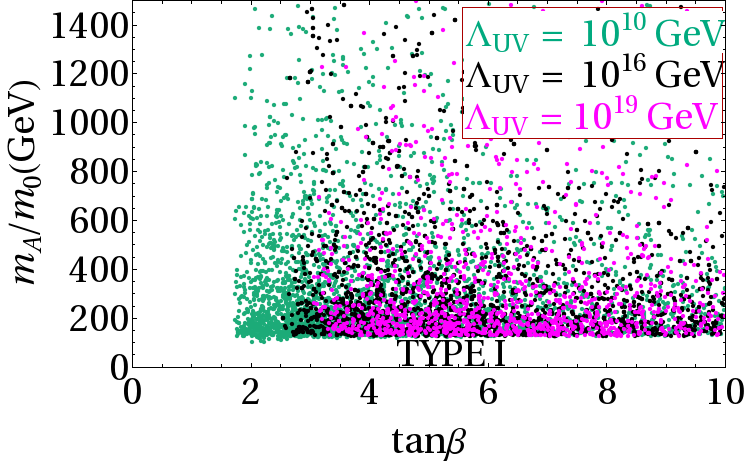} & 
			\includegraphics[width=5.2cm,height=4cm]{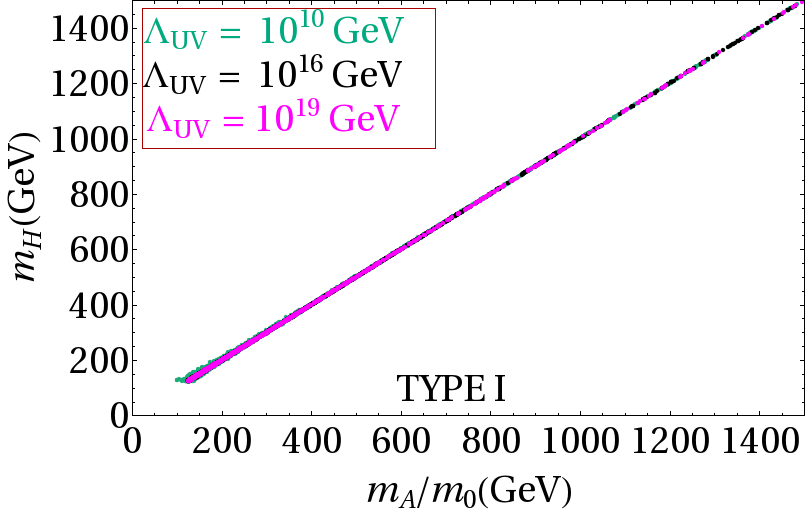} &
			\includegraphics[width=5.2cm,height=4cm]{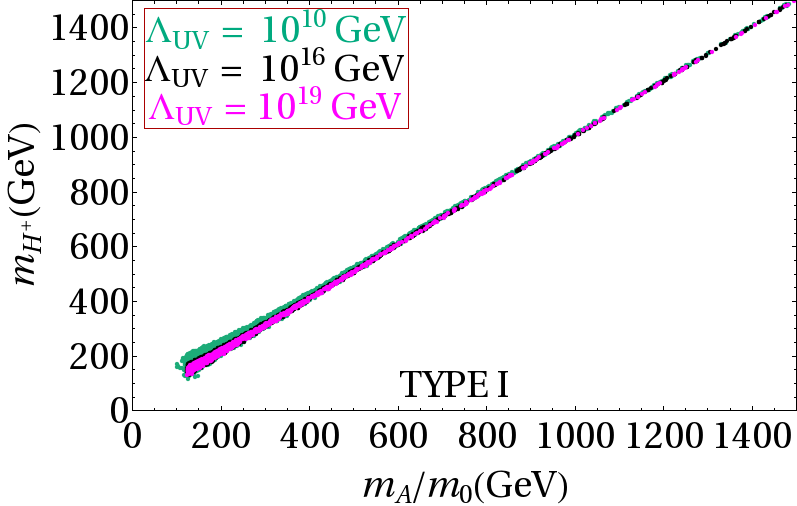}  
	\end{tabular}
	\caption{\em (Softly broken $U(1)$) Allowed points in $\tan\beta$-$m_A$, $m_A$-$m_H$, $m_A$-$m_{H^+}$ planes for the 2HDM potential to be stable up to $\luv$.
Three different colors correspond to three different choices of $\luv$ which appear in the legends.}
\label{broken_u1_mass}
\end{figure}

Here we will have $m_A=m_0$ from the beginning. Evidently, when the $U(1)$ symmetry is exact in the 
potential, {\em i.e.}, $m_0=0$, $A$ will become the Goldstone boson. We present the result of our analysis,
for Type~I models, in Fig.~\ref{broken_u1_mass}. Although these plots do not provide any new insights,
we include them just for completeness. The interpretation of Fig.~\ref{broken_u1_mass} will be similar to
that of the corresponding plots in the case of softly broken $Z_2$ symmetry.

\section{Summary and conclusions}\label{conclusions}
In this paper we have studied the consequences of demanding absolute stability of the 2HDM potential all the way
up to the Planck scale. In view of the fact that the LHC Higgs data are in conformity with the SM expectations, we
decided to work in the {\em alignment limit} where the lightest CP-even scalar ($h$) possess SM-like tree-level
couplings with the {\em standard} particles. Although we have explicitly demonstrated our results for Type~I
and II models only, they are equally applicable to Type~X and Y models also. We have found that, in the
alignment limit, requirement of high scale stability puts some remarkable restrictions on the 2HDM parameter
space. This set of restrictions defines the {\em stable alignment limit} for 2HDMs. Some of our most important observations are summarized below:
\begin{itemize}
\item 2HDM scalar potential with an exact $Z_2$ symmetry is unable to maintain stability after $10^8$~GeV
($10^4$~GeV) in the Type~I (II) case. To ensure stability up to the Planck scale, $Z_2$ symmetry needs
to be broken softly.
\item By demanding high scale stability in the presence of a soft breaking, we are led to a situation where the
symmetry of the potential is enhanced from softly broken $Z_2$ to softly broken $U(1)$.
\item To have stability up to very high energies ($\gtrsim 10^{10}$~GeV), all the nonstandard masses need to be nearly
degenerate: $m_0\approx m_A\approx m_H\approx m_{H^+}$. Thus, there is only one nonstandard mass
parameter that governs the 2HDM in the {\em stable alignment limit}.
\item The value of $\tb$ is bounded from below.
\end{itemize}
Therefore, in the {\em stable alignment limit}, a 2HDM can be fully described by only two new parameters,
namely, $\tb$ and $m_0$ with $\tb \gtrsim 3$ and $m_0\gtrsim 120$~GeV ($280$~GeV) for Type~I (II) models.

Finally, we end the paper on a pessimistic note. One should remember that, in the alignment limit, even the self couplings
of $h$ are identical to the SM at the tree-level. This means, in this limit, any measurement involving the 
{\em standard} particles can smell the presence of nonstandard physics only through loop effects. But then
the nonstandard particles can be made heavy enough to dilute these effects sufficiently. Thus, our only
hope to detect the presence of nonstandard scalars beyond the reaches of the direct search experiments
is via some nondecoupling effects in certain loop induced processes like Higgs to diphoton decay \cite{Bhattacharyya:2014oka}.
But with the degeneracy, $m_0^2\approx m_{H^+}^2$, this possibility is also  eliminated.
Therefore, it appears that even if the LHC data continues to agree with the SM predictions, a 2HDM in the
{\em stable alignment limit} which is very difficult, if not impossible, to probe experimentally, can serve as
an alternative to the SM for many years to come.
\paragraph*{Acknowledgments:}
DD thanks the Department of Atomic Energy (DAE), India for financial support.


\appendix
\section*{Appendix}
\section{One loop RG equations} 
\label{RGE_all}
Let us first write down the one loop RG equation (RGE) of the SM quartic coupling $(\lambda)$ \cite{Arason:1991ic,Luo:2002ey} as follows:
\begin{eqnarray}
{\cal D}\lambda &=& 12\lambda^2 -3\left(3g^2+{g^\prime}^2\right)\lambda + \frac{3}{4}\left(3g^4 + {g^\prime}^4 + 2g^2{g^\prime}^2\right) 
+12h_t^2\lambda - 12h_t^4 \,,
\label{SM_RG}
\end{eqnarray}
where, $h_t$ denotes the top-Yukawa which dominates over the other Yukawa couplings. For convenience,
we have introduce the shorthand ${\cal D}\equiv 16\pi^2\frac{\rm d}{\rm {d(ln\mu)}}$.

Now we will present the one loop RGEs 
of all the relevant couplings (gauge, Yukawa and scalar quartic couplings) of 2HDM \cite{Haber:1993an,Grimus:2004yh,Ferreira:2010xe,Branco:2011iw}. 
\paragraph{Gauge couplings:}
RGE for the gauge couplings, 
\begin{subequations}
	\begin{eqnarray}
	{\cal D}g_s &=& -7g_s^3\,,  \label{gs_rg}\\ 
	{\cal D}g &=& -3g^3\,,   \label{g_rg}\\
	{\cal D}g^\prime &=& 7{g^\prime}^3\,.  \label{gp_rg}
	\end{eqnarray}
	\label{gauge_rg}
\end{subequations}
\paragraph{Yukawa couplings:}
The initial values of top $(h_t)$, bottom $(h_b)$ and tau $(h_\tau)$ Yukawa couplings at
electroweak scale $(M_t)$ for the different Yukawa structures are given by,
\begin{subequations}
	\begin{eqnarray}
	{\rm Type~ I:} \left\{	
	\begin{array}{cll}
	h_{t}(M_t) &=& \frac{\sqrt{2}M_t}{v\sin\beta}\{1-\frac{4}{3\pi}\alpha_s(M_t)\}\,, \\
	h_{b,\tau}(M_t) &=& \frac{\sqrt{2}m_{b,\tau}}{v\sin\beta}\,;
	\end{array}
	\right.
	\end{eqnarray}
	\begin{eqnarray}
	{\rm Type~ II:} \left\{	
	\begin{array}{cll}
	h_{t}(M_t) &=& \frac{\sqrt{2}M_{t}}{v\sin\beta}\{1-\frac{4}{3\pi}\alpha_s(M_t)\}\,, \\
	h_{b,\tau}(M_t)  &=& \frac{\sqrt{2}m_{b,\tau}}{v\cos\beta}\,;
	\end{array}
	\right.
	\end{eqnarray}
	\label{yukawa_initial_values}
\end{subequations}
	where, $\alpha_s(M_t)$ denotes the strong coupling constant at top quark pole mass.
The corresponding RGEs for Type~I Yukawa structure are given by
\begin{subequations}
\begin{eqnarray}
{\cal D}h_t &=& h_t\left(a_u + \frac{9}{2}h_t^2 + \frac{3}{2}h_b^2 + h_{\tau}^2\right)\,,  \label{top_rg_t1}\\ 
{\cal D}h_b &=& h_b\left(a_d + \frac{3}{2}h_t^2 + \frac{9}{2}h_b^2 + h_{\tau}^2\right)\,,  \label{bot_rg_t1}\\ 
{\cal D}h_{\tau} &=& h_{\tau}\left(a_e + 3h_t^2 + 3h_b^2 + \frac{5}{2}h_{\tau}^2\right)\,. \label{tau_rg_t1}
\end{eqnarray} 
\end{subequations}
For Type~II Yukawa structure, the RGEs take the following form:
\begin{subequations}
\begin{eqnarray}
{\cal D}h_t &=& h_t\left(a_u + \frac{9}{2}h_t^2 + \frac{1}{2}h_b^2 \right)\,, \label{top_rg_t2} \\ 
{\cal D}h_b &=& h_b\left(a_d + \frac{1}{2}h_t^2 + \frac{9}{2}h_b^2 + h_{\tau}^2\right)\,,  \label{bot_rg_t2}\\ 
{\cal D}h_{\tau} &=& h_{\tau}\left(a_e + 3h_b^2 + \frac{5}{2}h_{\tau}^2\right)\,, \label{tau_rg_t2}
\end{eqnarray}
\end{subequations}
where,
\begin{subequations}
\begin{eqnarray}
a_u &=& \left(-8g_s^2 - \frac{9}{4}g^2 - \frac{17}{12}{g^\prime}^2\right)\,, \\ 
a_d &=& \left(-8g_s^2 - \frac{9}{4}g^2 - \frac{5}{12}{g^\prime}^2\right)\,, \\ 
a_e &=& \left(- \frac{9}{4}g^2 - \frac{15}{4}{g^\prime}^2\right)\,,
\end{eqnarray}
\label{yukawa_rg}
\end{subequations}
for both Type~I and II models.
\paragraph{Scalar quartic couplings:}
The RGEs for the five quartic couplings that appear in \Eqn{potential_notation1} are given by
\begin{eqnarray}
{\cal D}\lambda_i = \beta_{\lambda_i} + G_i + H_i\,, (i=1,2,3,4,5)\,,
\label{lambda_rg}
\end{eqnarray}
where, $\beta_{\lambda_i}$ and $G_i$ are independent of the Yukawa
structure of the model and are as follows:
\begin{subequations}
\begin{eqnarray}
\beta_{\lambda_1} &=& 12\lambda_1^2 + 4\lambda_3^2 + 4\lambda_3\lambda_4 + 2\lambda_4^2 + 2\lambda_5^2\,, \label{l1_rg_l}\\
\beta_{\lambda_2} &=& 12\lambda_2^2 + 4\lambda_3^2 + 4\lambda_3\lambda_4 + 2\lambda_4^2 + 2\lambda_5^2\,, \label{l2_rg_l}\\
\beta_{\lambda_3} &=& \left(\lambda_1+\lambda_2\right)\left(6\lambda_3 + 2\lambda_4\right) + 4\lambda_3^2 + 2\lambda_4^2 + 2\lambda_5^2\,, \label{l3_rg_l}\\
\beta_{\lambda_4} &=& 2\lambda_4\left(\lambda_1+\lambda_2\right) + 8\lambda_3\lambda_4 + 4\lambda_4^2 + 8\lambda_5^2\,, \label{l4_rg_l}\\
\beta_{\lambda_5} &=& \lambda_5\left(2\lambda_1 + 2\lambda_2 + 8\lambda_3 + 12\lambda_4\right)\,, \label{l5_rg_l} 
\end{eqnarray}
\label{lambda_rg_l}
\end{subequations}
and,
\begin{subequations}
\begin{eqnarray}
G_1 &=& \frac{3}{4}\left(3g^4 + {g^\prime}^4 + 2g^2{g^\prime}^2\right) - 3\lambda_1\left(3g^2 + {g^\prime}^2\right)\,, \label{l1_rg_g}\\
G_2 &=& \frac{3}{4}\left(3g^4 + {g^\prime}^4 + 2g^2{g^\prime}^2\right) - 3\lambda_2\left(3g^2 + {g^\prime}^2\right)\,, \label{l2_rg_g}\\
G_3 &=& \frac{3}{4}\left(3g^4 + {g^\prime}^4 - 2g^2{g^\prime}^2\right) - 3\lambda_3\left(3g^2 + {g^\prime}^2\right)\,, \label{l3_rg_g}\\
G_4 &=& 3g^2{g^\prime}^2 - 3\lambda_4\left(3g^2 + {g^\prime}^2\right)\,, \label{l4_rg_g}\\
G_5 &=& - 3\lambda_5\left(3g^2 + {g^\prime}^2\right)\,.  \label{l5_rg_g}
\end{eqnarray}
\label{lambda_rg_g}
\end{subequations}
The expressions for $H_i$s, however, depend on the Yukawa structure of the model. For Type~I models, these are given by,
\begin{subequations}
\begin{eqnarray}
H_1 &=& 0\,, \label{l1_yk_t1}\\
H_2 &=& 4\lambda_2\left(3h_t^2 + 3h_b^2 + h_{\tau}^2\right) - \left(12h_t^4 + 12h_b^4 + 4h_{\tau}^4\right)\,, \label{l2_yk_t1}\\
H_3 &=& 2\lambda_3\left(3h_t^2 + 3h_b^2 + h_{\tau}^2\right)\,, \label{l3_yk_t1}\\
H_4 &=& 2\lambda_4\left(3h_t^2 + 3h_b^2 + h_{\tau}^2\right)\,, \label{l4_yk_t1}\\
H_5 &=& 2\lambda_5\left(3h_t^2 + 3h_b^2 + h_{\tau}^2\right)\,.  \label{l5_yk_t1} 
\end{eqnarray}
\label{lambda_yk_t1}
\end{subequations}
For the Type~II Yukawa structure, we have
\begin{subequations}
\begin{eqnarray}
H_1 &=& 4\lambda_1\left(3h_b^2 + h_{\tau}^2\right) - \left(12h_b^4 + 4h_{\tau}^4\right)\,, \label{l1_yk_t2}\\
H_2 &=& 12\lambda_2h_t^2 - 12h_t^4 \,, \label{l2_yk_t2}\\
H_3 &=& 2\lambda_3\left(3h_t^2 + 3h_b^2 + h_{\tau}^2\right) - 12h_t^2h_b^2\,, \label{l3_yk_t2}\\
H_4 &=& 2\lambda_4\left(3h_t^2 + 3h_b^2 + h_{\tau}^2\right) + 12h_t^2h_b^2\,, \label{l4_yk_t2}\\
H_5 &=& 2\lambda_5\left(3h_t^2 + 3h_b^2 + h_{\tau}^2\right)\,.  \label{l5_yk_t2}
\end{eqnarray}
\label{lambda_yk_t2}
\end{subequations}

\bibliographystyle{JHEP}
\bibliography{reference.bib}
\end{document}